%% file: www2023.tex
\documentclass[sigconf]{acmart}

\usepackage{algorithm}
\usepackage{algorithmic}
\usepackage{multicol}
\usepackage{multirow}
\usepackage{subfigure}
\usepackage{tikz}
\usepackage{amsmath}
\usepackage{enumitem}
\usepackage{booktabs}
\usepackage{bbding}
\usepackage{graphicx}
\usepackage{stfloats}
\usepackage{amsfonts}
\usepackage{babel}

\newcommand{\CorrectSign}{\includegraphics[scale=0.22]{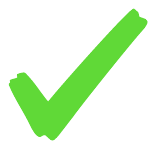}}
\newcommand{\WrongSign}{\includegraphics[scale=0.2]{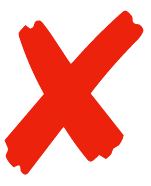}}
\newcommand{\SkipSign}{\includegraphics[scale=0.22]{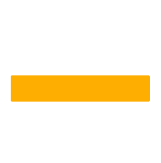}}

\makeatletter
\newcommand\footnoteref[1]{\protected@xdef\@thefnmark{\ref{#1}}\@footnotemark}
\makeatother

\AtBeginDocument{%
  }

\setcopyright{acmcopyright}
\copyrightyear{2023} 
\acmYear{2023} 
\setcopyright{acmlicensed}\acmConference[WWW '23]{Proceedings of the ACM Web Conference 2023}{May 1--5, 2023}{Austin, TX, USA}
\acmBooktitle{Proceedings of the ACM Web Conference 2023 (WWW '23), May 1--5, 2023, Austin, TX, USA}
\acmPrice{15.00}
\acmDOI{10.1145/3543507.3583866}
\acmISBN{978-1-4503-9416-1/23/04}

\begin{document}
\begin{sloppypar}

\title{Enhancing Deep Knowledge Tracing with Auxiliary Tasks}


\author{Zitao Liu}
\affiliation{%
  \institution{Guangdong Institute of Smart Education\\Jinan University}
  \city{Guangzhou}
  \country{China}
}
\email{liuzitao@jnu.edu.cn}

\author{Qiongqiong Liu}\authornote{The corresponding author: Qiongqiong Liu.}
\affiliation{%
  \institution{Think Academy International Education\\TAL Education Group}
  \city{Beijing}
  \country{China}
}
\email{liuqiongqiong1@tal.com}

\author{Jiahao Chen}
\affiliation{%
  \institution{Think Academy International Education\\TAL Education Group}
  \city{Beijing}
  \country{China}
}
\email{chenjiahao@tal.com}

\author{Shuyan Huang}
\affiliation{%
  \institution{Think Academy International Education\\TAL Education Group}
  \city{Beijing}
  \country{China}
}
\email{huangshuyan@tal.com}

\author{Boyu Gao}
\affiliation{%
  \institution{Guangdong Institute of Smart Education\\Jinan University}
  \city{Guangzhou}
  \country{China}
}
\email{bygao@jnu.edu.cn}

\author{Weiqi Luo}
\affiliation{%
  \institution{Guangdong Institute of Smart Education\\Jinan University}
  \city{Guangzhou}
  \country{China}
}
\email{lwq@jnu.edu.cn}

\author{Jian Weng}
\affiliation{%
  \institution{College of Information Science and Technology\\Jinan University}
  \city{Guangzhou}
  \country{China}
}
\email{cryptjweng@gmail.com }







\renewcommand{\shortauthors}{Liu et al.}

\begin{abstract}

Knowledge tracing (KT) is the problem of predicting students' future performance based on their historical interactions with intelligent tutoring systems. Recent studies have applied multiple types of deep neural networks to solve the KT problem. However, there are two important factors in real-world educational data that are not well represented. First, most existing works augment input representations with the co-occurrence matrix of questions and knowledge components\footnote{\label{ft:kc}A KC is a generalization of everyday terms like concept, principle, fact, or skill.} (KCs) but fail to explicitly integrate such intrinsic relations into the final response prediction task. Second, the individualized historical performance of students has not been well captured. In this paper, we proposed \emph{AT-DKT} to improve the prediction performance of the original deep knowledge tracing model with two auxiliary learning tasks, i.e., \emph{question tagging (QT) prediction task} and \emph{individualized prior knowledge (IK) prediction task}. Specifically, the QT task helps learn better question representations by predicting whether questions contain specific KCs. The IK task captures students' global historical performance by progressively predicting student-level prior knowledge that is hidden in students' historical learning interactions. We conduct comprehensive experiments on three real-world educational datasets and compare the proposed approach to both deep sequential KT models and non-sequential models. Experimental results show that \emph{AT-DKT} outperforms all sequential models with more than 0.9\% improvements of AUC for all datasets, and is almost the second best compared to non-sequential models. Furthermore, we conduct both ablation studies and quantitative analysis to show the effectiveness of auxiliary tasks and the superior prediction outcomes of \emph{AT-DKT}. To encourage reproducible research, we make our data and code publicly available at \url{https://github.com/pykt-team/pykt-toolkit}\footnote{We merged our model to the \textsc{pyKT} benchmark at \url{https://pykt.org/}.}.
\end{abstract}




\maketitle

\section{Introduction}
\label{sec:intro}
\input{intro}

\section{Related Work}
\label{sec:bg_mt}
\input{bg_mt}

\section{Problem Statement}
\label{sec:ps}
\input{ps}

\section{Enhancing Deep Knowledge Tracing with Auxiliary Tasks}
\label{sec:method}
\input{bg_dkt.tex}

\input{method}

\section{Experiment}
\label{sec:exp}
\input{exp.tex}


\section{Conclusions}
\label{sec:conclusion}
\input{conclusion.tex}



\begin{acks}
This work was supported in part by National Key R\&D Program of China, under Grant No. 2022YFC3303600; in part by Beijing Nova Program (Z201100006820068) from Beijing Municipal Science \& Technology Commission; in part by NFSC under Grant No. 61877029; in part by Key Laboratory of Smart Education of Guangdong Higher Education Institutes, Jinan University (2022LSYS003) and in part by National Joint Engineering Research Center of Network Security Detection and Protection Technology.
\end{acks}

\bibliographystyle{ACM-Reference-Format}
\bibliography{www2023}





\newpage
\onecolumn
\appendix
\input{appendix.tex}

\end{sloppypar}
\end{document}

%% file: intro.tex
\begin{figure*}[!bpht]
    \begin{minipage}[!bpht]{0.98\textwidth}
        \centering
        \includegraphics[width=\textwidth]{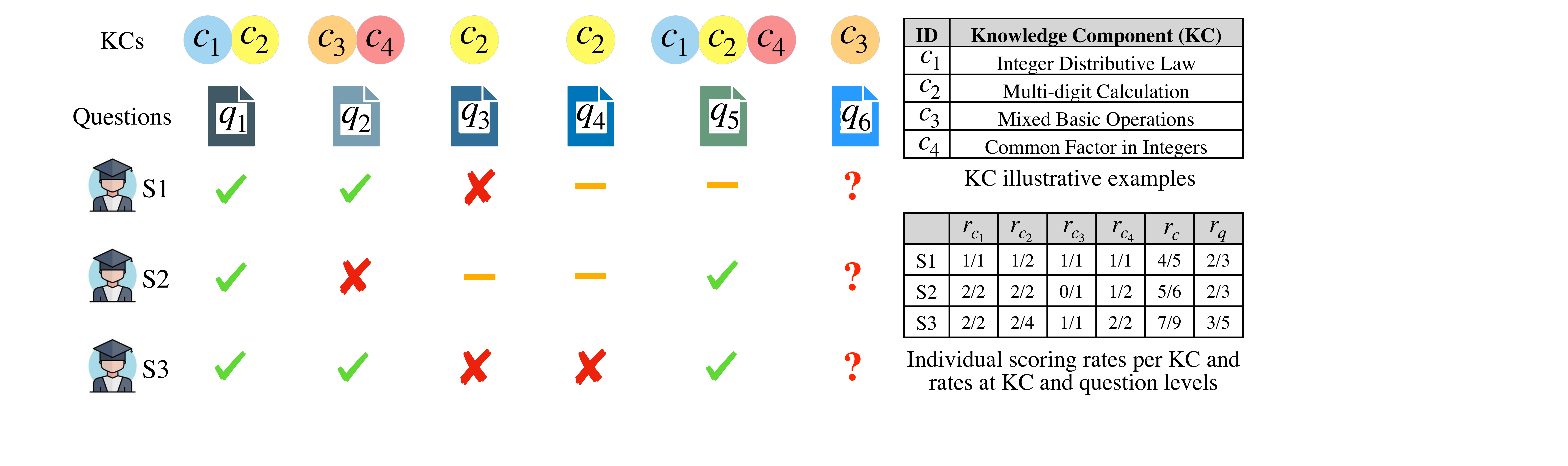} 
        \caption{KT illustration and a toy example of the individualization effect. $r_{c_i}$, $r_c$ and $r_q$ represent the individualized historical scoring rates for KC $c_i$, all KCs, and all questions respectively.``\CorrectSign'' and ``\WrongSign'' denote the question is answered correctly and incorrectly and ``\SkipSign'' denotes the student doesn't get chance to answer the question.}\label{fig:kt_illustration}
    \end{minipage}
    \Description[KT illustration]{KT illustration}
\end{figure*}

Knowledge tracing (KT) is a sequential prediction task that aims to predict the outcomes of students over questions by modeling their mastery of knowledge, i.e., knowledge states, as they interact with learning platforms such as massive open online courses and intelligent tutoring systems, as shown in Figure \ref{fig:kt_illustration}. Solving the KT problems may help teachers better detect students that need further attention, or recommend personalized learning materials to students.

The KT related research has been studied since the 1990s when Corbett and Anderson, to the best of our knowledge, were the first to estimate students' current knowledge with regard to each individual knowledge component (KC) \citep{corbett1994knowledge}. A KC is a description of a mental structure or process that a learner uses, alone or in combination with other KCs, to accomplish steps in a task or a problem\footnoteref{ft:kc}. Since then, many attempts have been made to solve the KT problem, such as probabilistic graphical models \citep{kaser2017dynamic} and factor analysis based models \citep{cen2006learning,lavoue2018adaptive,thai2012factorization}. 

Recently, with the rapid development of deep neural networks, many deep learning based knowledge tracing (DLKT) models are developed, such as auto-regressive based deep sequential KT models \cite{piech2015deep, yeung2018addressing,chen2018prerequisite,guo2021enhancing, long2021tracing,chen2023improving}, memory-augmented KT models \cite{zhang2017dynamic,abdelrahman2019knowledge,yeung2019deep}, attention based KT models \cite{pandey2019self,pandey2020rkt,choi2020towards,ghosh2020context,pu2020deep}, and graph based KT models \cite{nakagawa2019graph,yang2020gikt,tong2020hgkt}. Besides model variations in terms of neural architectures, a large spectrum of DLKT models are designed to incorporate as much as possible learning related information to augment its prediction ability. Such supplemental information includes question texts \cite{liu2019ekt,wang2020neural}, question similarities \cite{liu2021improving,wang2019deep}, question difficulties \cite{ghosh2020context,liu2021improving,shen2022assessing}, and relations between questions and KCs \cite{tong2020hgkt,pandey2020rkt,liu2021improving,yang2020gikt}.

Although the aforementioned DLKT approaches have constituted new paradigms of the KT problem and achieved promising results, two important factors in real-world educational data are not well represented. First, existing explorations of modeling the intrinsic relations between questions and KCs and building accurate student answer predictors are loosely connected. Previous approaches tend to learn relation enhanced embeddings from graphs involving questions, KCs and students and then augment the initial model input with the learned representations \cite{tong2020hgkt,pandey2020rkt,liu2021improving,yang2020gikt}. Unfortunately, such a graph is extremely sparse in real-world data. For example, Table \ref{tab:data_stats} shows basic data statistics of three widely used KT benchmark datasets. The majority of questions are only associated with 1 or 2 KCs and the average numbers of KC per question are 1.3634, 1.0136, and 1.0148 for the above datasets. Furthermore, due to the fact that such associations are manually annotated, mislabeled relations are inevitable and the corresponding errors might be easily propagated in the learning process of graph based DLKT models \cite{wang2020neural}.

\begin{table}
\caption{Data statistics of three widely used KT datasets containing both question and KC information. The details of the AL2005, BD2006, NIPS34 datasets are described in Section \ref{sec:dataset}. ``avg/p50/p90/p95/p99 KCs per Qs'' denote the average/median/90th/95th/99th percentile numbers of KCs per question.}
\label{tab:data_stats}
\begin{tabular}{llll}
\toprule
& \textbf{AL2005}  & \textbf{BD2006}    & \textbf{NIPS34}    \\
\midrule
\# of interactions   & 607,021 & 1,817,458 & 1,382,678 \\
\# of students       & 574     & 1,145     & 4,918     \\
\# of questions      & 173,113 & 129,263   & 948       \\
\# of KCs            & 112     & 493       & 57        \\ \hline
avg KCs per Qs  	 & 1.3634  & 1.0136    & 1.0148    \\ 
p50 KCs per Qs  	 & 1       & 1         & 1         \\
p90 KCs per Qs  	 & 2       & 1         & 1         \\  
p95 KCs per Qs 		 & 2       & 1         & 1         \\
p99 KCs per Qs 		 & 3       & 2         & 2         \\
\bottomrule
\end{tabular}
\end{table}

Second, many existing DLKT models assess knowledge states without explicitly capturing the student-level variability, i.e., \emph{individualization}, such as different knowledge acquisition abilities and learning rates. Modeling such student-level individualization could benefit the KT model's statistical goodness of fit, as well as potentially improve the generalization of the KT model \cite{yudelson2016individualizing,pardos2010modeling,shen2020convolutional}. Figure \ref{fig:kt_illustration} shows a toy example that illustrates the individualization effects on the KT prediction tasks, where 3 students have answered 5 questions related to 4 KCs. As we can see from Figure \ref{fig:kt_illustration}, even though student $S_1$ and student $S_2$ have the same historical scoring rate at the question level\footnote{The scoring rate is a fraction of the number of times one student correctly answers questions or KCs divided by the total number of times that the student practices such questions or KCs.}, their knowledge mastery levels (eg. scoring rate per KC.) differ a lot. Meanwhile, student $S_1$ and student $S_3$ have the same scoring rate per KC, but their question-level and KC-level historical ratings are quite different. Unfortunately, such individualization information of different students is not given in advance, which makes it very challenging to measure them.

In this paper we develop solutions that are applicable and can learn KT models from real-world educational contexts. Our work focuses on the refinements of a popular DLKT model: the Deep Knowledge Tracing (DKT) \cite{piech2015deep} and its application to student assessment. We aim to develop an algorithm to automatically learn a DKT that performs better student assessment by addressing the aforementioned two challenges.

Briefly, the DKT is one of the most widely used models of using deep neural networks to capture student interaction dynamics in the KT domain \cite{yeung2018addressing,chen2018prerequisite,guo2021enhancing, long2021tracing}. This is due to its relative simplicity, mathematically accurate prediction behavior, and the fact that it still leads the leaderboard across 7 popular datasets across different education domains according to a recent DLKT benchmarking report \cite{liu2022pykt}. The DKT is Markovian and assumes the dynamic knowledge states of the student are captured well using a small set of real-valued hidden-state variables. The DKT can be learned from observational interaction data with any gradient descent optimization algorithms.

In this work, we address the above issues by introducing two auxiliary learning tasks including: 

\begin{itemize}
\item \emph{Question tagging (QT) prediction task}: automatically predicting whether questions contain specific KCs.

\item \emph{Individualized prior knowledge (IK) prediction task}: progressively predicting student-level prior knowledge that is hidden in students' historical learning interactions.
\end{itemize}

Our approach builds upon the original auto-regressive DKT architecture, and augments its original cross-entropy objective function that optimizes the probabilities that a student can correctly answer questions with two auxiliary tasks. In the QT task, we use a Transformer encoder with a masked attention mechanism to extract contextual similar question-level information that is relevant to the exercise to be answered and assign KCs to each question. In the IK task, at each time step, we use a student ability network to measure individualized historical performance considering all the previous questions and responses for each student. To ensure that our approach can be fairly comparable with other recently developed DLKT models, we choose to follow a publicly available standardized KT task evaluation protocol \cite{liu2022pykt}. We conduct rigorous experiments on three public datasets and the results show that our auxiliary task enhanced DKT model, i.e., \emph{AT-DKT}, is able to improve the ``simple but tough-to-beat'' DKT model in terms of AUC by at least 0.9\%.


%% file: bg_mt.tex
A large spectrum of approaches has been developed to enhance the original objective function that predicts students' next interaction performance via explicitly adding either regularization penalties or auxiliary learning tasks\footnote{Broadly speaking, the regularization penalty terms can also be viewed as a part of auxiliary tasks.}. For example, \citet{sonkar2020qdkt} used a Laplacian regularizer to incorporate the assumption that the success probabilities of multiple questions are associated with the same KC should not be significantly different for a given learner. \citet{chen2018prerequisite} proposed to restrict the model learning under the prerequisite KC ordering pair constraints. \citet{yeung2018addressing} developed three regularization terms to address the reconstruction and wavy transition problems in the original DKT model. \citet{zhang2021multi} added regularization terms about question difficulty level to restrict question representations to fine-tuning. \citet{wang2019deep} designed an auxiliary task to model the hierarchical relations between KCs and questions. \citet{guo2021enhancing} constructed adversarial examples to improve the generalization of the original DKT model with extra adversarial perturbation loss. For KT tasks on multiple choice questions, \citet{an2022no} proposed a multi-task learning framework to predict both the student's correctness and option choice for a given question.

%% file: ps.tex
Our objective is given an arbitrary question $q_*$ to predict the probability of whether a student will answer $q_*$ correctly or not based on the student's historical interaction data. More specifically, for each student $\mathbf{S}$, we assume that we have observed a chronologically ordered collection of $T$ past interactions i.e., $\mathbf{S} = \{\mathbf{s}_j\}_{j=1}^T$. Each interaction is represented as a 4-tuple $\mathbf{s}$, i.e., $\mathbf{s} = <q, \{c\}, r, t>$, where $q, \{c\}, r, t$ represent the specific question, the associated KC set, the binary valued student response, and student's response time step respectively. The response is a binary valued indicator variable where 1 represents the student correctly answered the question, and 0 otherwise. We would like to estimate the probability $\hat{r}_{*}$ of the student's performance on the arbitrary question $q_*$.

%% file: bg_dkt.tex
\subsection{Notation}

In real-world educational scenarios, the question bank is usually much bigger than the set of KCs. For example, the number of questions is more than 1500 times larger than the number of KCs in the AL2005 dataset\footnote{Details about the AL2005 dataset is described in Section \ref{sec:dataset}.}. Following the recommended suggestions by \cite{piech2015deep,ghosh2020context,liu2022pykt}, we train the DLKT models on KC-response data, which is artificially generated from question-response data by expanding each question-level interaction into multiple KC-level interactions when the question is associated with a set of KCs. Notations are based on this expanded KC-response data in the rest of the paper. 

Let $\mathbf{X} = \{\mathbf{x}_t\}_{t=1}^{T'}$ be the expanded student's historical interaction sequence at the KC level, where $T'$ is the length of the interaction sequence. Let $\mathbf{x}_t$, $\mathbf{q}_t$ and $\mathbf{c}_t$ represent the $d$-dimensional dense interaction, question, KC embeddings at time step \textit{t}, i.e., $\mathbf{x}_t, \mathbf{q}_t, \mathbf{c}_t \in \mathbb{R}^{d \times 1}$. Let $M$ and $N$ be the total number of questions and KCs. Let $\oplus$, $\mathbf{I}(\cdot)$ $ReLU(\cdot)$ and $\sigma(\cdot)$ be the element-wise addition operation, indicator function, rectified linear unit activation function and the element-wise sigmoid function.

\subsection{Deep Knowledge Tracing}
\label{sec:bg_dkt}

The DKT model predicts students' future performance by using an auto-regressive neural architecture to capture the evolving changes of students' knowledge states \cite{piech2015deep}. In DKT, each interaction $\mathbf{x}_t$ is generated by a response encoder that encodes both the KC and response related information at time step $t$. Then, it uses a recurrent neural network to obtain a $d$-dimensional hidden state $\mathbf{h}_t$ to represent the current knowledge state and uses a linear feed forward network layer to output the estimated levels of knowledge mastery $\hat{\mathbf{r}}_t$, i.e.,

\vspace{-0.3cm}
\begin{align*}
\mathbf{x}_t & = \mathbf{W}_0 \mathbf{e}_t; \mathbf{h}_t  = \mbox{LSTM}(\mathbf{h}_{t-1}, \mathbf{x}_t);
\hat{\mathbf{r}}_t = \sigma (\mathbf{W}_1 \mathbf{h}_t + \mathbf{b}_1)
\end{align*}

\noindent where $\mathbf{W}_0$, $\mathbf{W}_1$ and $\mathbf{b}_1$ are learnable parameters and $\mathbf{W}_0 \in \mathbb{R}^{2N \times d}$, $\mathbf{W}_1 \in \mathbb{R}^{N \times d}$, $\mathbf{b}_1 \in \mathbb{R}^{N \times 1}$. $\mathbf{e}_t$ is the one-hot encoding of the raw input representation, i.e., $\mathbf{e}_t \in \{0, 1\}^{2N}$, and $e_t^k = 1$ if the answer is wrong and $e_t^{k+N} = 1$ if the answer is correct. $k$ represents the KC index at time step $t$.


%% file: method.tex
\subsection{The AT-DKT Framework}

\begin{figure}[!bpht]
\centering
\vspace{-0.4cm}
\includegraphics[width=0.9\columnwidth]{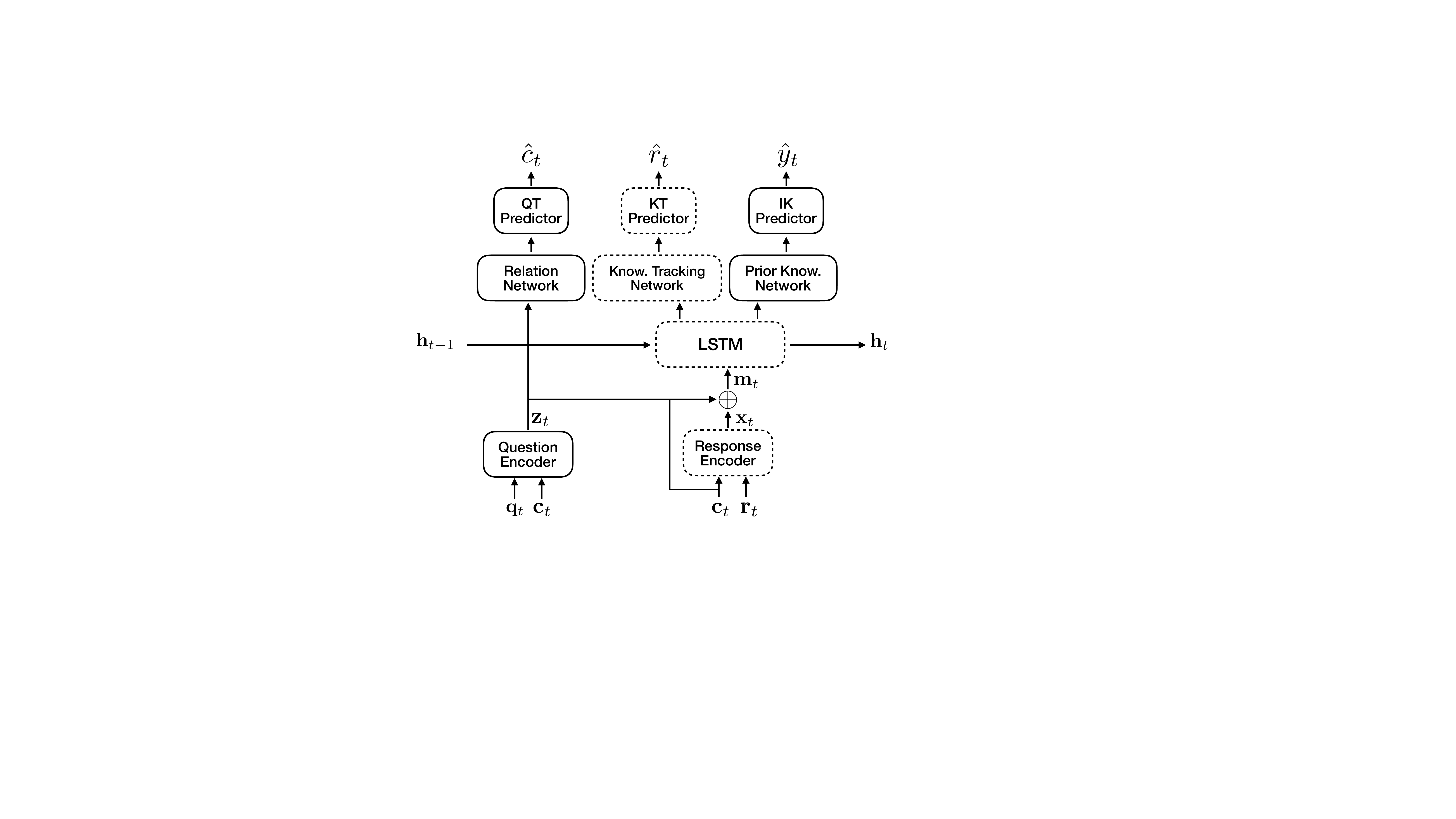}\vspace{-0.2cm}
\caption{The overview of the proposed AT-DKT framework.}
\vspace{-0.5cm}
\label{fig:model}
\Description[AT-DKT framework]{AT-DKT framework}
\end{figure}

In this section, we present the  \textit{AT-DKT} framework overview (shown in Figure \ref{fig:model}) that improves the original DKT model with two auxiliary learning tasks. In Figure \ref{fig:model}, we outline the key components of the original DKT model with dash lines and visualize the auxiliary  tasks enhanced components in AT-DKT with solid lines. In the AT-DKT framework, we explicitly incorporate the additional QT and IK prediction tasks together with the original KT predictor to better student assessment. The QT task focuses on predicting the assigned KCs to the questions by modeling the intrinsic relations among the questions and KCs with students' previous learning outcomes and the IK task aims to estimate the individualized historical performance of each student according to their learning processes.

\subsection{Question Tagging Prediction Task}
\label{sec:at1}

\input{at1.tex}

\subsection{Individualized Prior Knowledge Prediction Task}
\label{sec:at2}
\input{at2.tex}

\subsection{Knowledge Tracing Predictor}
\label{sec:emb}

Our AT-DKT is built upon the original DKT model and chooses to use the auto-regressive neural architecture as the backbone of our AT-DKT model. In addition, we make two modifications to improve the modeling ability of the standard DKT model. First, different from the DKT model that only relies on the representations of KCs and responses, we add question specific information $\mathbf{z}_t$ and $\mathbf{c}_t$ into the interaction encoding, i.e., $\mathbf{m}_t = \mathbf{z}_t \oplus \mathbf{c}_t \oplus \mathbf{x}_t$. This will help the deep sequential based model to capture more question-sensitive information. Second, the vanilla version of DKT makes the KT predictions from the linearly transformed hidden knowledge states while in our AT-DKT model, we replace such linear transformation operation with a non-linear projection of two-layer neural networks, i.e., 

\vspace{-0.2cm}
\begin{align*}
\hat{\mathbf{r}}_{t} = \sigma \Bigl(\mathbf{W}^r_2 \cdot \mbox{ReLU} ( \mathbf{W}^r_1 \cdot \mathbf{h}_t + \mathbf{b}^r_1 ) + \mathbf{b}^r_2 \Bigl)
\end{align*}

\noindent where $\mathbf{W}^r_1$, $\mathbf{W}^r_2$, $\mathbf{b}^r_1$ and $\mathbf{b}^r_2$ are trainable parameters and $\mathbf{W}^r_1 \in \mathbb{R}^{d/2 \times d}$, $\mathbf{W}^r_2 \in \mathbb{R}^{N \times d/2}$, $\mathbf{b}^r_1 \in \mathbb{R}^{d/2 \times 1}$, $\mathbf{b}^r_2 \in \mathbb{R}^{N \times 1}$.

The knowledge tracing predictor is optimized by minimizing the binary cross-entropy loss between the ground-truth response $r_{t+1}$ and the prediction probability $\hat r_{t}$ of the KC $c_{t+1}$. Let $\delta (c_{t+1})$ be the one-hot encoding of the KC is answered at time $t+1$, the loss of this task is:

\vspace{-0.2cm}
\begin{align*}
	\mathcal{L}_{\mbox{KT}} = \sum -\left(r_{t+1} \log \hat{\mathbf{r}}_{t} \delta (c_{t+1})+\left(1-r_{t+1}\right) \log \left(1-\hat{\mathbf{r}}_{t} \delta (c_{t+1})\right)\right)
\end{align*}

\subsection{Model Optimization}
\label{sec:opt}

All parameters in the entire \textit{AT-DKT} model are optimized together in a unified framework by minimizing the three losses above. The final loss of our method is:

\vspace{-0.2cm}
\begin{align*}
\mathcal{L} = \mathcal{L}_{\mbox{KT}} + \beta_1 \mathcal{L}_{\mbox{QT}} + \beta_2 \mathcal{L}_{\mbox{IK}}   \\
\end{align*}
\vspace{-0.5cm}

\noindent where $\beta_1$, $\beta_2$ are the tuning hyper-parameters.

%% file: at1.tex
Different from previous graph based DLKT approaches that first learn representations of questions and KCs from their relation graph and then concatenate the learned embeddings as part of the model input, we improve the representation discriminative ability by explicitly using the intermediate representations to predict whether a KC is associated with the question at each time step. Specifically, we construct the initial question representation $\mathbf{a}_t$ by combining both the question and KC embedding, i.e., $\mathbf{a}_t = \mathbf{q}_t + \mathbf{c}_t$. Then, based on each student's learning history, we use a Transformer based question encoder with a masked dot production attention function to capture the long-term contextual dependencies of both questions and KCs within the student learning history. This operation not only models the temporal recency effects during student learning \cite{zhang2021multi} but tries to find latent connections among questions and KCs. 

To assess the relevance between current representation $\mathbf{a}_t$ at time step $t$ and previously answered questions $\mathbf{a}_i$, we compute the relevant coefficients $\alpha_t^i$ by taking the Softmax activation of the masked dot product between $\mathbf{a}_t$ and $\mathbf{a}_i$, i.e., $\alpha_t^i = \mbox{Softmax}( \mathbf{a}_i \cdot \mathbf{a}_t  ), i = 1, \cdots, t$. These relevant coefficients are used as attention scores in the standard multi-head Transformer encoders to obtain the enhanced question representation $\mathbf{z}_t$, i.e., $\mathbf{z}_t =\mbox{MultiHeadTransformer}(\{\mathbf{a}_i\}_{i=1}^t)$.

To further conduct the QT task, we design a relation network first to extract the representation of the question and KC with a fully connected neural layer and then project it into the KC space via non-linear transformation. We use the output of the relation network for the QT prediction. The predicted result of KC membership for the specific question at time step $t$ is computed as follows:

\vspace{-0.1cm}
\begin{align*} 
\hat{\mathbf{c}}_{t} = \sigma \Bigl(\mathbf{W}^c_2 \cdot \mbox{ReLU} ( \mathbf{W}^c_1 \cdot \mathbf{z}_t + \mathbf{b}^c_1 ) + \mathbf{b}^c_2 \Bigl)
\end{align*}

\noindent where $\mathbf{W}^c_1$, $\mathbf{W}^c_2$, $\mathbf{b}^c_1$ and $\mathbf{b}^c_2$ are trainable parameters and $\mathbf{W}^c_1 \in \mathbb{R}^{d/2 \times d}$, $\mathbf{W}^c_2  \in \mathbb{R}^{N \times d/2}$, $\mathbf{b}^c_1 \in \mathbb{R}^{d/2 \times 1}$, $\mathbf{b}^c_2 \in \mathbb{R}^{N \times 1}$.

The QT prediction task is optimized by minimizing the binary cross-entropy loss between the ground-truth KC tag $c_{t}$ and the prediction result $\hat{c}_{t}$ of the question $q_{t}$:

\vspace{-0.1cm}
\begin{equation}
\label{equ:at1_loss}
	\mathcal{L}_{\mbox{QT}} = \sum -\left(c_{t} \log \hat{c}_{t}+\left(1-c_{t}\right) \log \left(1-\hat{c}_{t}\right)\right)  \nonumber
\end{equation}

%% file: at2.tex
In real-world educational contexts, generally speaking, students' prior knowledge is embedded in their historical learning interactions \cite{shen2020convolutional,zhang2021multi}. Several research works have demonstrated that the scoring rates of students could be viewed as the reflection of their overall knowledge mastery \cite{wang2020neural}. Therefore, we explicitly add the IK task to measure the individualized prior knowledge of students comprehensively based on their historical learning interactions. The student-level scoring rate accounts for the overall knowledge mastery on all the KCs learned so far and is computed as $y_{t} = ({\sum_{j=1}^{t} \mathbf I(r_{j}==1})) / {t}$. Please note that we choose to not use the individualized KC-wise scoring rate because the student interaction sequence is usually very short and the frequency based KC-wise scoring rate becomes very unreliable.


In this work, we introduce a time-aware student prior knowledge estimation task to progressively predict student overall scoring rate. Specifically, once we have the current knowledge state $\mathbf{h}_t$ updated by the LSTM cell at each time step $t$, we use a prior knowledge network of a two-layer fully connected networks to summarize the students' mastery of prior knowledge.  The prior knowledge network will conduct non-linear transformations on the general knowledge states and convert it to the estimated historical scoring rate $\hat{y}_t$ as follows:

\vspace{-0.3cm}
\begin{align*}
\hat{\mathbf{y}}_{t} = \sigma \Bigl(\mathbf{W}^y_2 \cdot \mbox{ReLU} ( \mathbf{W}^y_1 \cdot \mathbf{h}_t + \mathbf{b}^y_1 ) + \mathbf{b}^y_2 \Bigl)
\end{align*}


\noindent where $\mathbf{h}_t = LSTM(\mathbf{h}_{t-1}, \mathbf{m}_t)$, $\mathbf{m}_t = \mathbf{z}_t \oplus \mathbf{c}_t \oplus \mathbf{x}_t$, $\mathbf{W}^y_1$, $\mathbf{W}^y_2$, $\mathbf{b}^y_1$ and $\mathbf{b}^y_2$ are trainable parameters and $\mathbf{W}^y_1 \in \mathbb{R}^{d/2 \times d}$, $\mathbf{W}^y_2 \in \mathbb{R}^{1 \times d/2}$, $\mathbf{b}^y_1 \in \mathbb{R}^{d/2 \times 1}$, $\mathbf{b}^y_2 \in \mathbb{R}^{1}$.



We use the mean square loss function to measure the accuracy of our estimated scoring rating in the IK task as follows:

\vspace{-0.3cm}
\begin{align*}
  \mathcal{L}_{\mbox{IK}} = \sum \mathbf{I}(t > \delta) (y_{t}-\hat{y_{t} })^{2}
\end{align*}

\noindent where $\delta$ is the tuning hyper-parameter that controls the length of historical observations and helps the model avoid noisy scoring rating calculation when $t$ is too small.

%% file: exp.tex

\subsection{Datasets}
\label{sec:dataset}
\input{dataset.tex}

\subsection{Baselines}
\label{sec:baselines}

\input{baselines.tex}

\begin{table*}[!bpht]
\small
\setlength\tabcolsep{10pt}
\centering
\caption{Performance comparisons in terms of AUC and accuracy for deep sequential models. Marker $*$, $\circ$ and $\bullet$ indicate whether AT-DKT is statistically superior/equal/inferior to the compared method (using paired t-test at 0.01 significance level).} \vspace{-0.3cm}
\label{tab:sequential}
\begin{tabular}{llllllll}
\hline
\multirow{2}{*}{Model} & \multicolumn{3}{c}{\textbf{AUC}}                       & \multirow{2}{*}{} & \multicolumn{3}{c}{\textbf{Accuracy}}                  \\ \cline{2-4} \cline{6-8} 
                       & \multicolumn{1}{c}{AL2005} & \multicolumn{1}{c}{BD2006} & \multicolumn{1}{c}{NIPS34} &                   & \multicolumn{1}{c}{AL2005} & \multicolumn{1}{c}{BD2006} & \multicolumn{1}{c}{NIPS34}         \\ \hline
\textbf{DKT}           & 0.8149±0.0011* & 0.8015±0.0008*       & 0.7689±0.0002* &                   & 0.8097±0.0005* & 0.8553±0.0002*       & 0.7032±0.0004* \\
\textbf{DKT+}          & 0.8156±0.0011* & 0.8020±0.0004*       & 0.7696±0.0002* &                   & 0.8097±0.0007* & 0.8553±0.0003*       & 0.7039±0.0004* \\
\textbf{DKT-F}         & 0.8147±0.0013* & 0.7985±0.0013*       & 0.7733±0.0003* &                   & 0.8090±0.0005* & 0.8536±0.0004*       & 0.7076±0.0002* \\
\textbf{KQN}           & 0.8027±0.0015* & 0.7936±0.0014*       & 0.7684±0.0003* &                   & 0.8025±0.0006* & 0.8532±0.0006*       & 0.7028±0.0001* \\
\textbf{ATKT}       & 0.7995±0.0023* & 0.7889±0.0008$\circ$ & 0.7665±0.0001* &                   & 0.7998±0.0019* & 0.8511±0.0004$\circ$ & 0.7013±0.0002* \\ \hline
\textbf{AT-DKT}        & 0.8246±0.0018  & 0.8105±0.0009        & 0.7816±0.0002  &                   & 0.8144±0.0009  & 0.8560±0.0005        & 0.7145±0.0002* \\ \hline
\end{tabular} \vspace{-0.2cm}
\end{table*}

\begin{table*}[!bpht]
\small
\setlength\tabcolsep{10pt}
\centering
\caption{Performance comparisons in terms of AUC and accuracy for non-sequential models. Marker $*$, $\circ$ and $\bullet$ indicate whether AT-DKT is statistically superior/equal/inferior to the compared method (using paired t-test at 0.01 significance level).} \vspace{-0.3cm}
\label{tab:non-sequential}
\begin{tabular}{llllllll}
\hline
\multirow{2}{*}{Model} & \multicolumn{3}{c}{\textbf{AUC}}                                       & \multirow{2}{*}{} & \multicolumn{3}{c}{\textbf{Accuracy}}                                  \\ \cline{2-4} \cline{6-8} 
                       & \multicolumn{1}{c}{AL2005} & \multicolumn{1}{c}{BD2006} & \multicolumn{1}{c}{NIPS34} &                   & \multicolumn{1}{c}{AL2005} & \multicolumn{1}{c}{BD2006} & \multicolumn{1}{c}{NIPS34} \\ \hline
\textbf{DKVMN}         & 0.8054±0.0011$\circ$ & 0.7983±0.0009*         & 0.7673±0.0004*         &                   & 0.8027±0.0007$\circ$ & 0.8545±0.0002*         & 0.7016±0.0005*         \\
\textbf{SKVMN} & 0.7463±0.0022* & 0.7310±0.0065* & 0.7513±0.0005* &  & 0.7837±0.0023* & 0.8404±0.0007* & 0.6885±0.0004* \\
\textbf{GKT}           & 0.8110±0.0009*       & 0.8046±0.0008*         & 0.7689±0.0024*         &                   & 0.8088±0.0008*       & 0.8555±0.0002*         & 0.7014±0.0028*         \\
\textbf{SAKT}          & 0.7880±0.0063*       & 0.7740±0.0008$\circ$   & 0.7517±0.0005*         &                   & 0.7954±0.0020*       & 0.8461±0.0005$\circ$   & 0.6879±0.0004*         \\
\textbf{SAINT}         & 0.7775±0.0017*       & 0.7781±0.0013*         & 0.7873±0.0007$\bullet$ &                   & 0.7791±0.0016*       & 0.8411±0.0065*         & 0.7180±0.0006$\bullet$ \\
\textbf{AKT}           & 0.8306±0.0019$\circ$ & 0.8208±0.0007$\bullet$ & 0.8033±0.0003$\bullet$ &                   & 0.8124±0.0011$\circ$ & 0.8587±0.0005$\bullet$ & 0.7323±0.0005$\bullet$ \\ \hline
\textbf{AT-DKT}        & 0.8246±0.0018        & 0.8105±0.0009          & 0.7816±0.0002          &                   & 0.8144±0.0009        & 0.8560±0.0005          & 0.7145±0.0002          \\ \hline
\end{tabular}
\vspace{-0.3cm}
\end{table*}

\vspace{-0.2cm}
\subsection{Experimental Setup}
\label{sec:setup}
\input{setup.tex}

\subsection{Results}
\label{sec:results}
\input{results.tex}

%% file: dataset.tex
We select three public real-world educational  datasets to evaluate the effectiveness of our model. 

\begin{itemize}[leftmargin=*]
	\item Algebra 2005-2006\footnote{\label{ft:l1}\quad\url{https://pslcdatashop.web.cmu.edu/KDDCup/}} (AL2005): This dataset stems from KDD Cup 2010 EDM Challenge which includes 13-14 year-old students' interactions with Algebra questions. It has detailed step-level student responses to the mathematical problems \cite{stamper2010algebra}. In our experiment, we use the concatenation of the problem name and step name as a unique question.
	\item Bridge to Algebra 2006-2007\footnoteref{ft:l1} (BD2006): Similar to AL2005, the data of BD2006 are mathematical problems from logs of students' interactions with intelligent tutoring systems \cite{stamper2010algebra}. The unique question construction of BD2006 is similar to AL2005.
	\item NeurIPS2020 Education Challenge\footnote{\quad\url{https://eedi.com/projects/neurips-education-challenge}} (NIPS34): This dataset is provided by NeurlPS 2020 Education Challenge. We use the dataset of Task 3 \& Task 4 to evaluate our models \cite{wang2020instructions}. The dataset contains students' answers to mathematics questions from Eedi which millions of students interact with daily around the globe.
\end{itemize}



To conduct reproducible experiments, we attentively follow the data pre-processing steps suggested in \cite{liu2022pykt}. We remove student sequences shorter than 3 attempts and the maximum length of student interaction history is set to 200 for high computational efficiency.

%% file: baselines.tex

\subsubsection{Deep Sequential KT Models}

Deep sequential KT models utilize an auto-regressive architecture to capture the intrinsic dependencies among students' chronologically ordered interactions \cite{chen2018prerequisite,guo2021enhancing,lee2019knowledge,liu2019ekt,minn2018deep,nagatani2019augmenting,piech2015deep,su2018exercise,yeung2018addressing}. Since the very first and successful research work of DKT that applies recurrent neural networks to model students' dynamic learning behaviors by \citet{piech2015deep}, a large number of works have been done to improve DKT's performance \cite{yeung2018addressing,chen2018prerequisite,su2018exercise,nagatani2019augmenting,lee2019knowledge,liu2019ekt,guo2021enhancing}. In this work, we select 5 widely used baselines as follows:

\begin{itemize}[leftmargin=*]
\item \textit{DKT} \cite{piech2015deep}: It uses Recurrent Neural Networks(RNNs) to model student learning and predict the mastery of each KC after one response to a new question. In this paper, we use LSTM as the base RNN cell. DKT is the first model which uses deep learning to KT.
\item \textit{DKT+} \cite{yeung2018addressing}: This method was proposed for solving the two problems of DKT, the first problem is that DKT fails to reconstruct the observed input, and the second problem is the inconsistent performance of KCs across time-steps. For the second problem, the authors used both \emph{L1}-norm and \emph{L2}-norm to measure the difference between two adjacent prediction results.
\item \textit{DKT-F} \cite{nagatani2019augmenting}: This model is also an extension of DKT, which adds a forgetting mechanism to predict the performance of users. The authors proposed three time-related features to improve the original DKT model, which are repeated time gap, sequence time gap and past trial counts.
\item \textit{KQN} \cite{lee2019knowledge}: It uses neural networks to encode the students' ability and skill vectors respectively and uses the dot product of the two types of vectors to do prediction. The authors introduced probabilistic skill similarity to make KQN interpretable and intuitive.
\item \textit{ATKT} \cite{guo2021enhancing}: It uses adversarial perturbations to enhance the KT model's generalization and reduce the overfitting problem of the DNNs-based KT models. The adversarial perturbations and the original interaction embedding are added to predict the students' performance. In this paper, an attention-LSTM was used as the KT backbone.
\end{itemize}

\subsubsection{Deep Non-Sequential KT Models}

Besides the deep sequential KT models, other types of neural network based approaches are applied in the KT domain as well, such as memory augmented KT models that explicitly model latent relations between KCs with an external memory \cite{abdelrahman2019knowledge,shen2021learning,zhang2017dynamic}, graph based KT models that capture interaction relations with graph neural networks \cite{nakagawa2019graph,tong2020structure,yang2020gikt}, and attention based KT models that use the attention mechanism and its variants to capture dependencies between interactions \cite{ghosh2020context,pandey2020rkt,pu2020deep,zhang2021multi}. In this work, we select 6 widely used baselines as follows:

\begin{itemize}[leftmargin=*]
\item \textit{DKVMN} \cite{zhang2017dynamic}: This is a memory augmented neural network that exploits the relationships between underlying concepts by a key matrix and uses a value matrix to represent the student's mastery of each KC at each time step.
\item \textit{SKVMN} \cite{abdelrahman2019knowledge}: It exploits a key-value memory network to enhance the representation capability of students' knowledge state and capture long-term dependencies of a student's learning sequence based on Hop-LSTM.
\item \textit{GKT} \cite{nakagawa2019graph}: Inspired by the potential graph structure of coursework, GKT casts the knowledge structure as a graph and reformulates the KT task as a time series node-level classification problem in GNN. The authors also proposed various implementations of the graph structure to overcome the lack of graph structure in many datasets.
\item \textit{SAKT} \cite{pandey2019self}: This method uses self-attention to identify the relevance between KCs to improve the lack of generalization to deal with sparse data in other models. SAKT uses question embeddings as query and uses the interaction embeddings as key and value.
\item \textit{SAINT} \cite{choi2020towards}: It uses a Transformer-based model for KT, the encoder applies self-attention to the sequence of exercises, and the decoder applies self-attention and masked encoder-decoder attention to the sequence of responses.
\item \textit{AKT} \cite{ghosh2020context}: This model applies a novel monotonic attention mechanism to connect the learners' future performance to their past responses, and uses a Rasch model to regularize the KC and question embeddings.
\end{itemize}

%% file: setup.tex
To evaluate the models' performance, we perform standard 5-fold cross-validation for all the combinations of pairs of method and dataset. We choose to use early stopping when the performance is not improved after 10 epochs. For each hyperparameter combination, we use the Adam optimizer to train the models up to 200 epochs. We adopt the Bayesian search method to find the best hyper-parameters for each fold. Specifically, the embedding size is searched from \{64, 256\}, the search space of the hidden size of the LSTM module is the same as the embedding size, the number of layers and the number of attention heads of the relation network are set to \{1, 2, 4\} and \{4, 8\}, the model's learning rate is set to \{1e-3, 1e-4, 1e-5\}, the hyper-parameters $\delta$, $\beta_1$ and $\beta_2$ are searched from \{0, 10, 30, 50, 70, 100, 120, 150\}, \{0.01, 0.1, 0.3, 0.5, 0.7, 1.0\}, and \{0.01, 0.1, 0.3, 0.5, 0.7, 1.0\} respectively. Similar to existing works \cite{piech2015deep,yeung2018addressing,liu2022pykt}, we use the area under the receiver operating characteristics curve (AUC) and accuracy to evaluate the KT prediction performance.

%% file: results.tex
\subsubsection{Performance on Deep Sequential Models}
\label{sec:res_seq}

Table \ref{tab:sequential} shows the performance of all sequential models in the three datasets. We report the average AUCs and accuracy and the standard deviations across 5 folds. From the table, we have the following observations: (1) AT-DKT significantly outperforms all deep sequential KT models and improves the AUC of the original DKT model by 0.97\% for AS2005, 0.90\% for BD2006 and 1.27\% for NIPS34. This indicates the effectiveness of the proposed auxiliary learning tasks; and (2) DKT is still a very strong baseline compared to its variants such as DKT+ and DKT-F. The performance of DKT-F on NIPS34 is slightly better rather than which on AL2005 and BD2006. We believe this is because the interval time in NIPS34 is 35,904 seconds, which is about 5 times longer than the numbers in AL2005 and BD2006 datasets. The DKT-F model is able to capture latent time related information from long-span intervals.

\begin{figure*}[!bpht]
\centering
\includegraphics[width=0.85\textwidth]{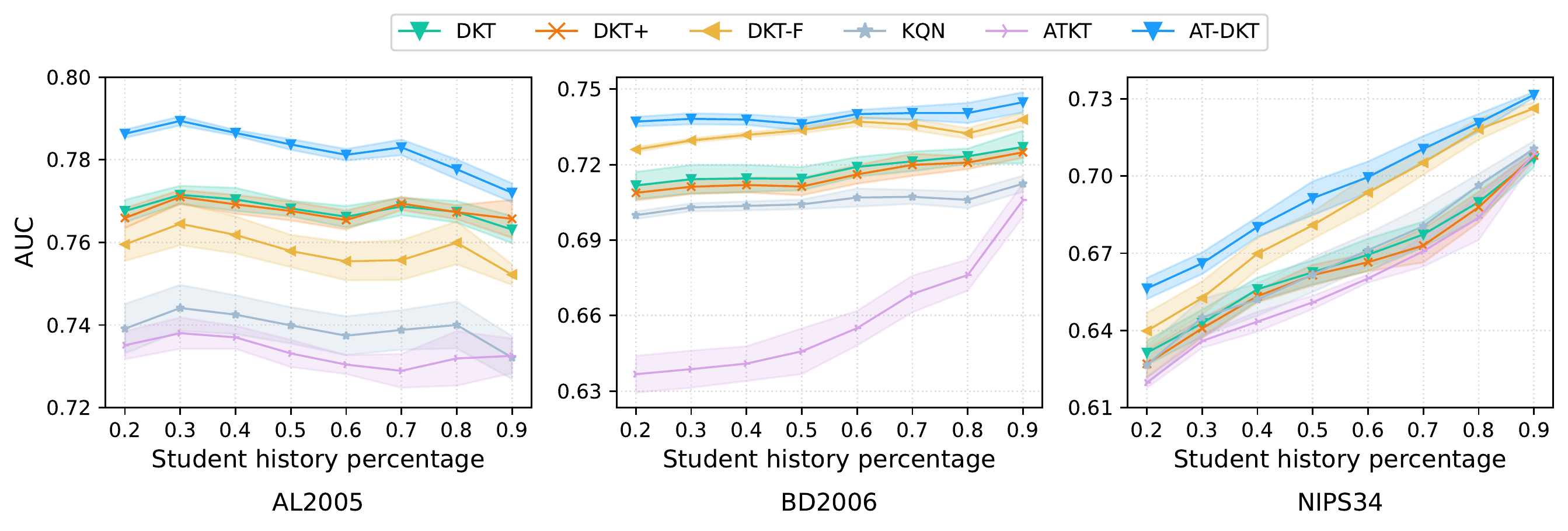}
\vspace{-0.4cm}
\caption{Accumulative predictions in the multi-step ahead scenario in terms of AUC on all datasets in sequential models.} \vspace{-0.1cm}
\label{fig:auc_accu_5}
\Description[AUC in sequential models]{AUC in sequential models}

\includegraphics[width=0.85\textwidth]{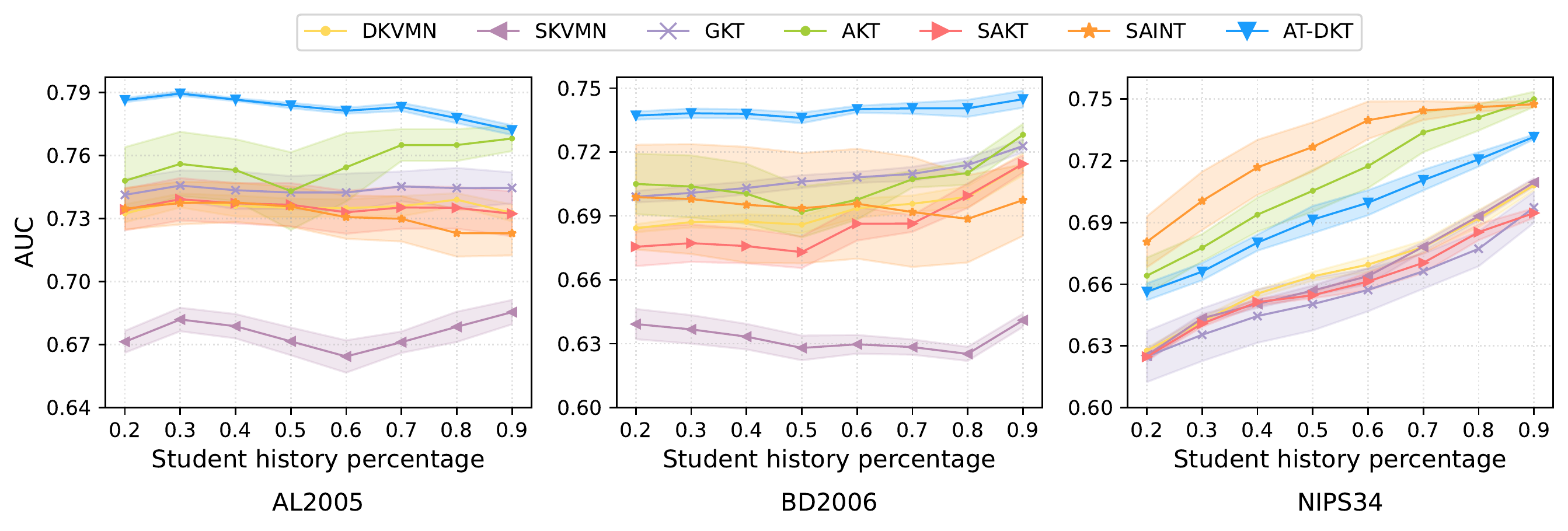}
\vspace{-0.3cm}
\caption{Accumulative predictions in the multi-step ahead scenario in terms of AUC on all datasets in non-sequential models.} \vspace{-0.3cm}
\label{fig:auc_accu_10}
\Description[AUC in non-sequential models]{AUC in non-sequential models}
\end{figure*}

\subsubsection{Performance on Deep non-Sequential Models}
\label{sec:res_non_seq}

We show the AUC and accuracy performance of the three datasets in non-sequential models in Table \ref{tab:non-sequential}. From the table, we can see that: (1) AT-DKT is on par with the best model on the AL2005 dataset, 1.03\% worse than AKT on the BD2006 dataset and 2.17\% and 0.57\% worse than AKT and SAINT on the NIPS34 dataset. We believe this is because the QT task in AT-DKT heavily relies on the relationships between KCs and questions. When comparing AL2005 to BD2006 and NIPS34, the AL2005 dataset has much denser question-KC associations and its averaged number of KCs per question is 1.3634, which is about 30\% larger than the numbers in BD2006 and NIPS34 datasets. The rich question-KC associations help AT-DKT achieves better representations for both questions and KCs; (2) AKT outperforms all the other non-sequential baseline models. We believe that this is because the AKT model utilizes not only a monotonic attention mechanism to capture short-term dependencies from past interactions at different time scales but also a Rasch model to consider the question difficulties; and (3) comparing DKVMN and SKVMN, SKVMN performs much worse than DKVMN. This is because the SKVMN only uses interactions of the same KCs and ignores interactions of similar KCs which is also very useful for student response prediction.

\subsubsection{Multi-step ahead KT Prediction}
\label{sec:res_multistep}

To make the KT prediction close to real application scenarios, we evaluate DLKT models in a multi-step prediction setting where the models are required to predict a future span of students' responses given their historical interactions. We conduct prediction in an accumulative prediction, i.e., use the model estimation of $\hat{r}_{t+1}, \hat{r}_{t+2}, \cdots, \hat{r}_{t+\Delta-1}$,  when predicting $\hat{r}_{t+\Delta}$ where $\Delta$ is an arbitrary future time span. Specifically, we vary the observed percentages of student interaction length from 20\% to 90\% with a step size of 10\%. In Figures \ref{fig:auc_accu_5} and \ref{fig:auc_accu_10}, we show the AUC performance of accumulative prediction in sequential and non-sequential models respectively. Due to the space limit, we provide the performance results in terms of accuracy in Appendix \ref{sec:app_acc}.

From Figures \ref{fig:auc_accu_5} and \ref{fig:auc_accu_10}, we can see that: (1) AT-DKT outperforms all other sequential models in accumulative prediction for all datasets, which indicates that the estimated responses made by AT-DKT are accurate and beneficial for the next-step prediction in sequential models; (2) compared to non-sequential methods, AT-DKT has the best performance on AL2005 and BD2006 datasets and is the second best model in NIPS34. We find the data in NIPS34 is multiple choice questions while the questions in AL2005 and BD2006 consist of step-level problems. Since the proposed auxiliary task IK is to measure the prior knowledge of students according to their historical learning interactions, the step-level problems are strongly correlated to get a better prediction result in the prior knowledge estimation task hence improving the student response prediction; (3) AT-DKT has a much smaller variance compared to other models indicating that our AT-DKT model is more stable and robust to varying prediction scenarios; and (4) with the increasing percentage of student historical interactions, the performance of all the DLKT models' gradually increase. This suggests that accurate student assessment requires  moderate-size student response observations.

\subsubsection{Visualization of Prediction Outcomes}
\label{sec:res_outcome}

\begin{figure*}[!bpht]
\centering
\includegraphics[width=0.9\textwidth]{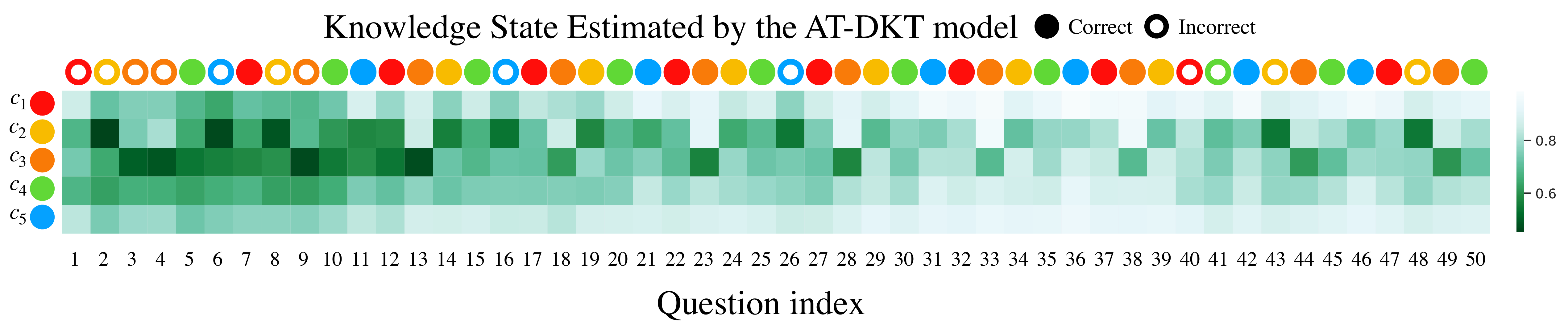} \vspace{-0.3cm}
\caption{An example of knowledge state changes of 5 concepts as a student solves 50 questions from the AL2005 dataset.} 
\label{fig:case}
\Description[case]{case} \vspace{-0.3cm}
\end{figure*}

The model estimated knowledge states of each concept will dynamically change when the student responds to each question. Therefore, in this section, we conduct qualitative analysis to investigate such a phenomenon. 
Figure \ref{fig:case} shows an example of changes in the estimated knowledge states of 5 concepts as a student solves 50 questions from the AL2005 dataset. 
As we can see, the estimated knowledge states of $c_2$/$c_5$ decrease/increase quite a bit when the student mistakenly/correctly answers the 2nd/11th question that contains the KC $c_2$/$c_5$ respectively. As the student practices more questions, the knowledge state estimations become much more stable and after finishing all these 50 questions, the model is confident that the student has acquired the KCs $c_1$ and $c_5$. Due to the space limit, we provide more visualization in Appendix \ref{sec:app_sample}.

\subsubsection{Visualization of $\mathbf{m}_{t}$}
\label{sec:res_vis}

\begin{figure*}[!bpht]
    \begin{minipage}[!bpht]{0.33\textwidth}
        \centering
        \includegraphics[width=\textwidth]{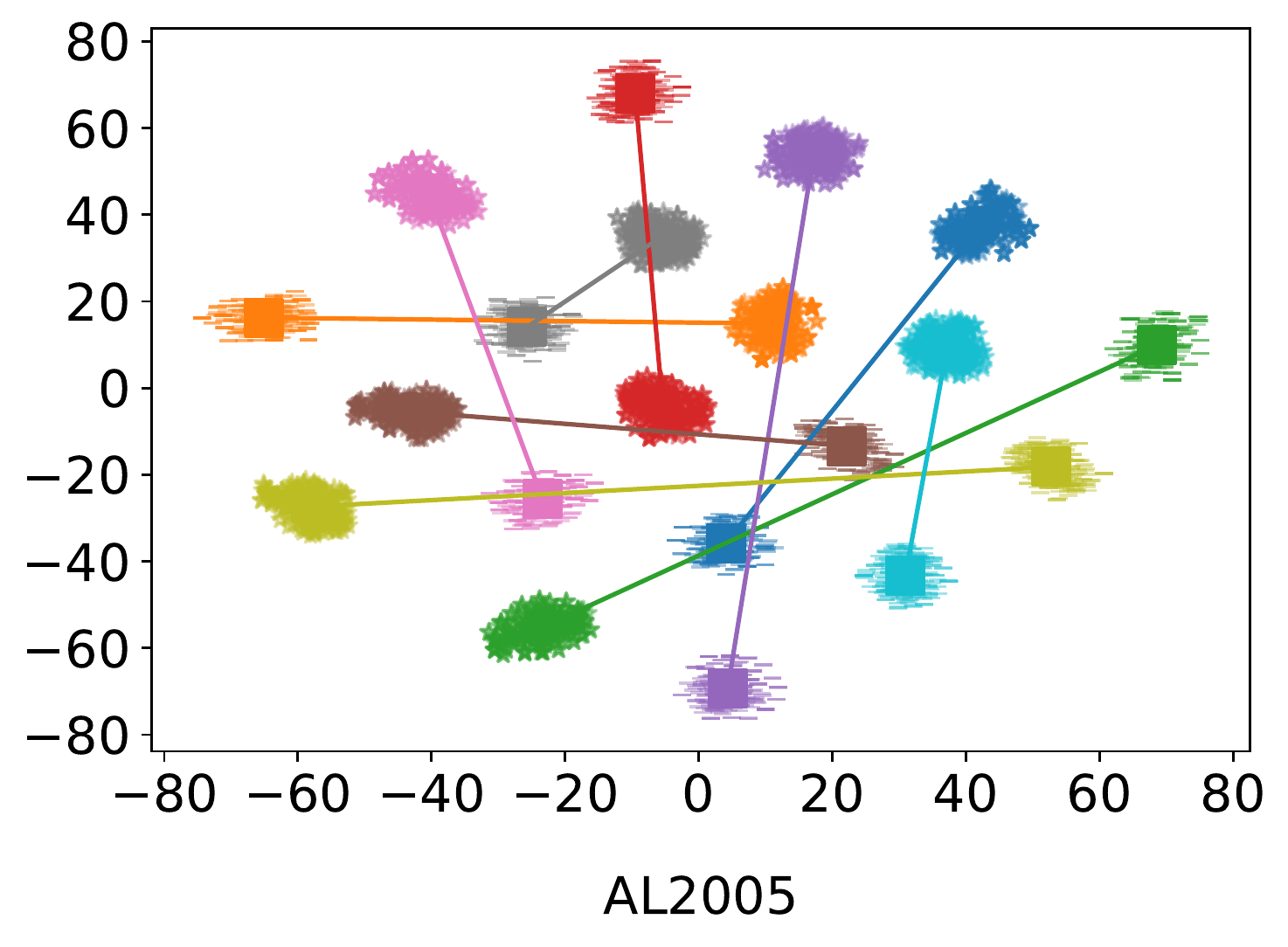} 
    \end{minipage}
    \begin{minipage}[!bpht]{0.33\textwidth}
        \centering
        \includegraphics[width=\textwidth]{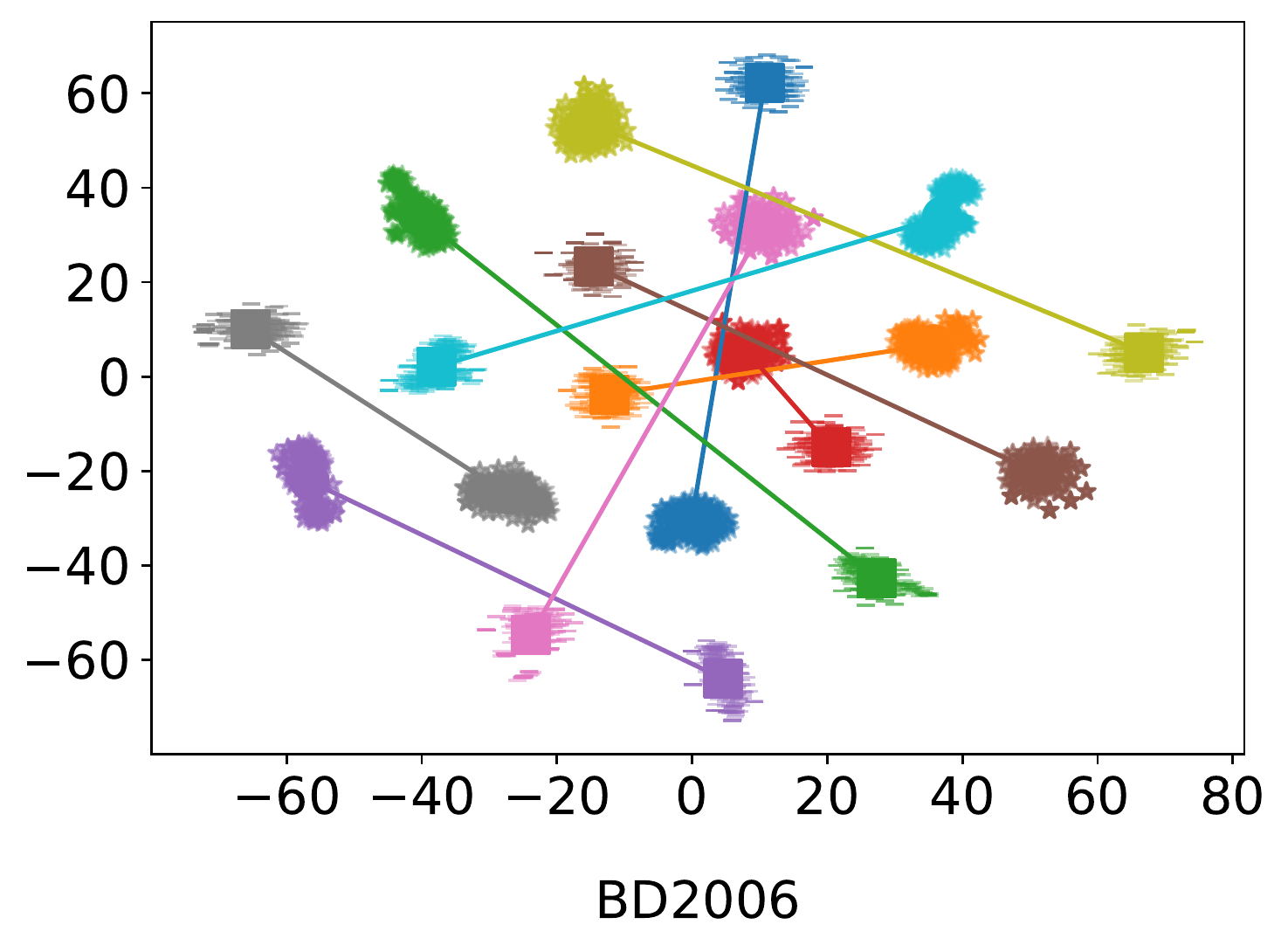} 
    \end{minipage}
    \begin{minipage}[!bpht]{0.33\textwidth}
        \centering
        \includegraphics[width=\textwidth]{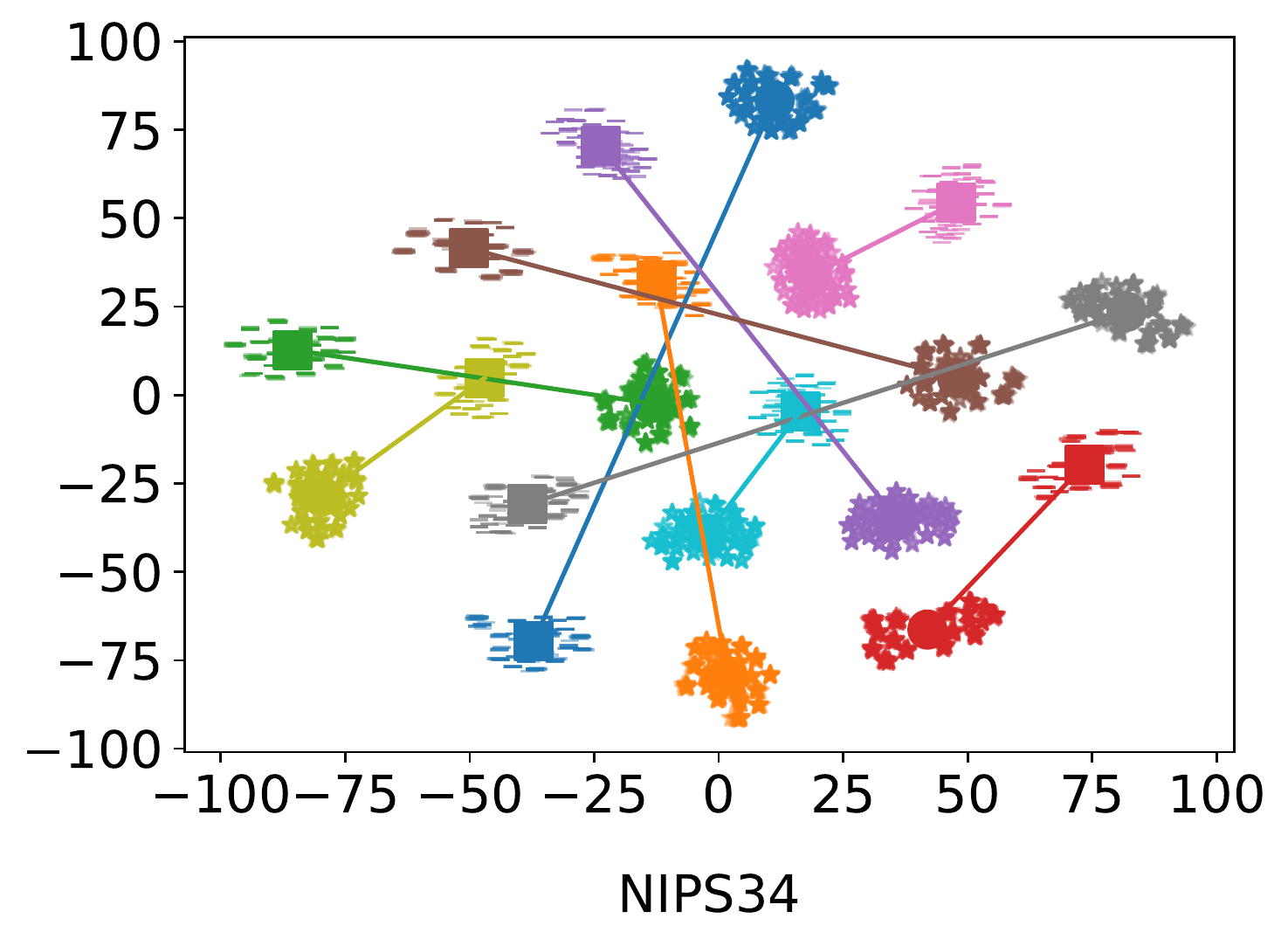} 
    \end{minipage} \vspace{-0.3cm}
     \caption{Visualization of $\mathbf{m}_{t}$ in AL2005, BD2006 and NIPS34 datasets. Best viewed zoomed in and in color.}\label{fig:mt} \vspace{-0.3cm}
    \Description[m_{t}]{m_{t}}
\end{figure*}

To examine the effectiveness of the learned representations, we conduct qualitative analysis on the combined representations $\mathbf{m}_{t}$s of questions, KCs and responses, i.e., $\mathbf{m}_t = \mathbf{z}_t \oplus \mathbf{c}_t \oplus \mathbf{x}_t$ (shown in Figure \ref{fig:model}). We first select the top 10 KCs from each dataset in terms of the number of student responses. For each selected KC, we randomly select 200 students' correct and incorrect responses respectively and the corresponding representation visualizations are shown in Figure \ref{fig:mt}. In Figure \ref{fig:mt}, different KCs are shown in different colors. The correct responses and incorrect responses are marked with horizontal lines and asterisks respectively. We draw a line between the centers of correct and incorrect representations for better visualization. From Figure \ref{fig:mt}, we can observe that the learned representation of $\mathbf{m}_{t}$ can well reconstruct the observed input. As a result, when a student performs well on a KC, the predicted mastery level of that KC increases. This alleviates the reconstruction problem mentioned in \cite{yeung2018addressing}.

\subsubsection{Ablation Study}
\label{sec:res_ablation}

We systematically examine the effect of key components by constructing four model variants in Table \ref{tab:ablation-dkt}. ``w/o'' means excludes such module from AT-DKT. Please note that AT-DKT w/o QT \& IK is equivalent to the vanilla DKT model proposed by \citep{piech2015deep}. From Table \ref{tab:ablation-dkt}, we can easily observe that (1) compared to other variants (e.g., AT-DKT w/o IK, AT-DKT w/o QT, and AT-DKT w/o IK \& QT), AT-DKT obtains the highest AUC score in all cases. This suggests that prediction performance degrades when ignoring either auxiliary learning task. Thus, it is important to incorporate such intrinsic information of question-KC relations and student-level prior knowledge in DLKT models; and (2) when comparing AT-DKT to AT-DKT w/o IK and AT-DKT w/o QT, we can see that the prediction improvements from the QT task and the IK task are complementary. The QT task is the leading contribution to the AUC score boost that the performance of AT-DKT drops 0.75\%, 0.85\% and 1.20\% for AL2005, BD2006 and NIPS34 datasets respectively without QT.

\begin{table}[!bpht]
\small
\setlength\tabcolsep{4pt}
\centering
\caption{Contribution analysis of the two auxiliary tasks.} 
\label{tab:ablation-dkt}
\begin{tabular}{lccc}
\hline
Models           & AL2005                 & BD2006                 & NIPS34                 \\ \hline
AT-DKT & \textbf{0.8246±0.0018} & \textbf{0.8105±0.0009} & \textbf{0.7816±0.0002} \\ 
AT-DKT w/o IK    & 0.8234±0.0016          & 0.8097±0.0011          & 0.7808±0.0005          \\
AT-DKT w/o QT    & 0.8171±0.0005          & 0.8020±0.0007          & 0.7696±0.0002          \\
AT-DKT w/o QT \& IK              & 0.8149±0.0011          & 0.8015±0.0008          & 0.7689±0.0002          \\ \hline
\end{tabular}
\vspace{-0.4cm}
\end{table}

%% file: conclusion.tex
In this paper, we propose to enhance the original deep knowledge tracing model with both the question tagging prediction task and the individualized prior knowledge prediction task. With these two auxiliary learning tasks, our AT-DKT approach is able to (1) directly mine intrinsic associations between questions and KCs and (2) extract students' global performance from their historical interactions. Experiment results on three real-world educational datasets demonstrated that AT-DKT outperforms a wide range of state-of-the-art DLKT learning approaches in terms of both AUC and accuracy. In the future, we plan to study a combination of our auxiliary learning tasks with non-sequential KT models. 

%% file: appendix.tex
\section{Performance results in terms of accuracy in the multi-step ahead scenario}
\label{sec:app_acc}

\begin{figure*}[!bpht]
\centering
\includegraphics[width=0.85\linewidth]{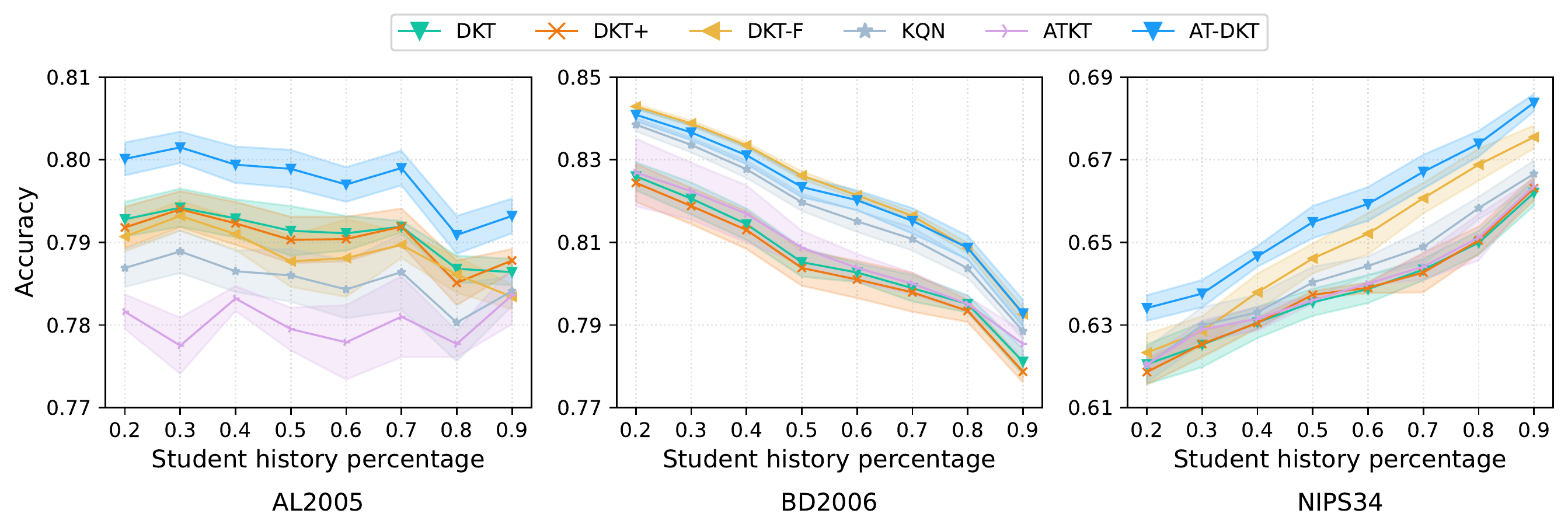} 
\caption{Accumulative predictions in the multi-step ahead scenario in terms of Accuracy on all datasets in sequential models.}
\label{fig:acc_accu_5}
\Description[Accuracy in sequential models]{Accuracy in sequential models}

\includegraphics[width=0.85\linewidth]{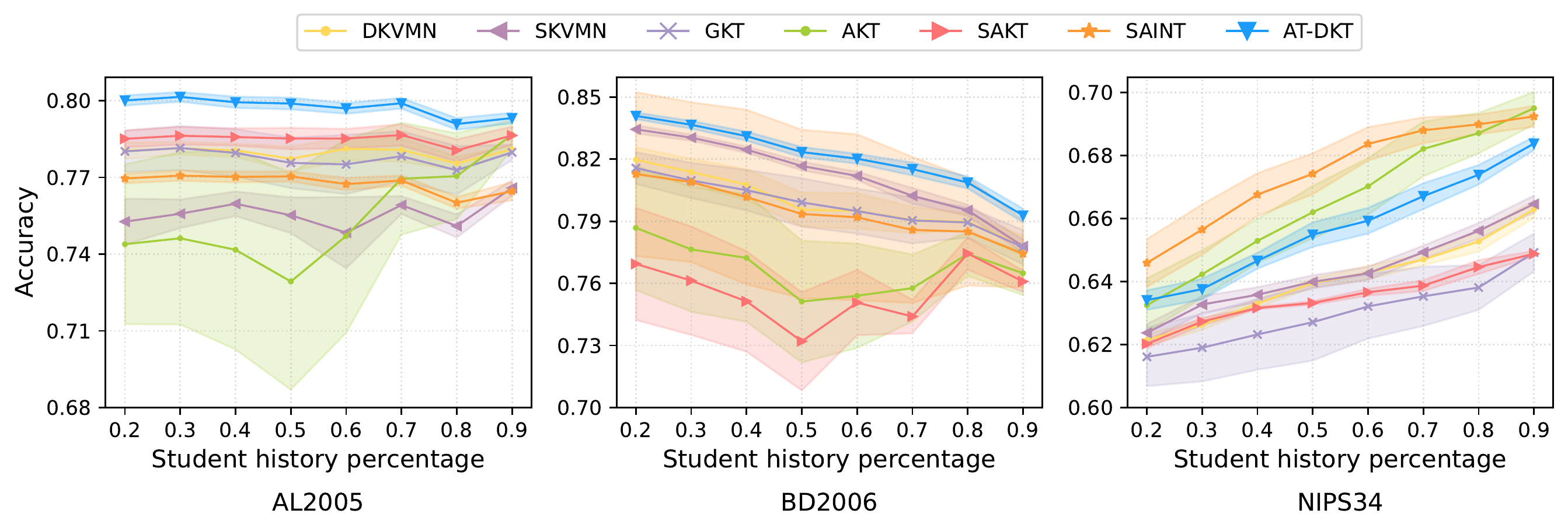}
\caption{Accumulative predictions in the multi-step ahead scenario in terms of Accuracy on all datasets in non-sequential models.}
\label{fig:acc_accu_10}
\Description[Accuracy in non-sequential models]{Accuracy in non-sequential models}
\end{figure*}

In Section \ref{sec:res_multistep}, we show the performance of AUC in accumulative prediction settings on all models. Here we show the accuracy of these model in Figure \ref{fig:acc_accu_5} and \ref{fig:acc_accu_10}.

\section{More visual samples of prediction outcomes}
\label{sec:app_sample}

\begin{figure*}[!bpht]
\centering
\includegraphics[width=0.9\textwidth]{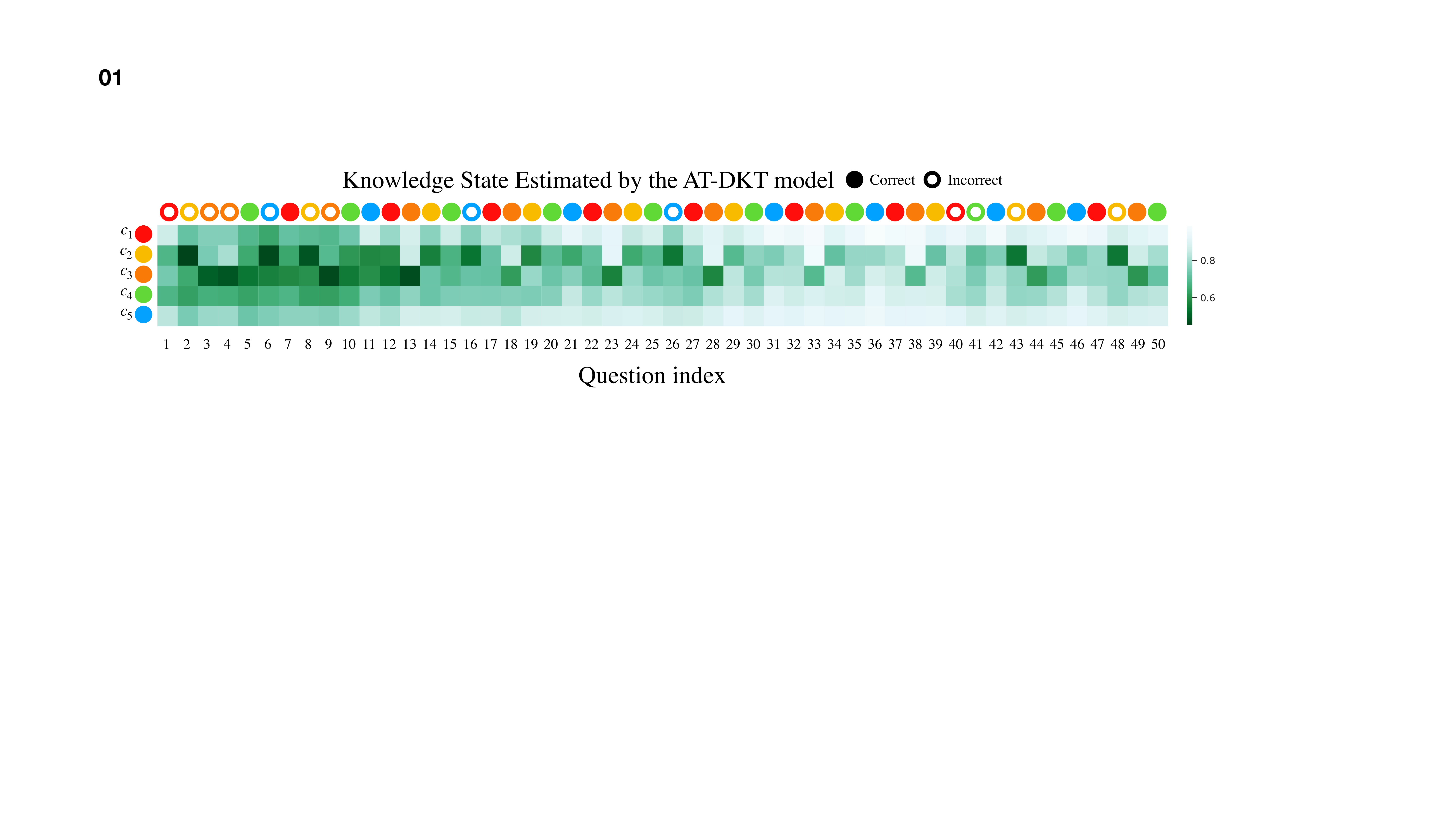} \vspace{-0.3cm}
\caption{An example of knowledge state changes of 5 concepts as the student 2 solves 50 questions from the AL2005 dataset.} 
\label{fig:case_2}
\Description[case]{case}
\end{figure*}

\begin{figure*}[!bpht]
\centering
\includegraphics[width=0.9\textwidth]{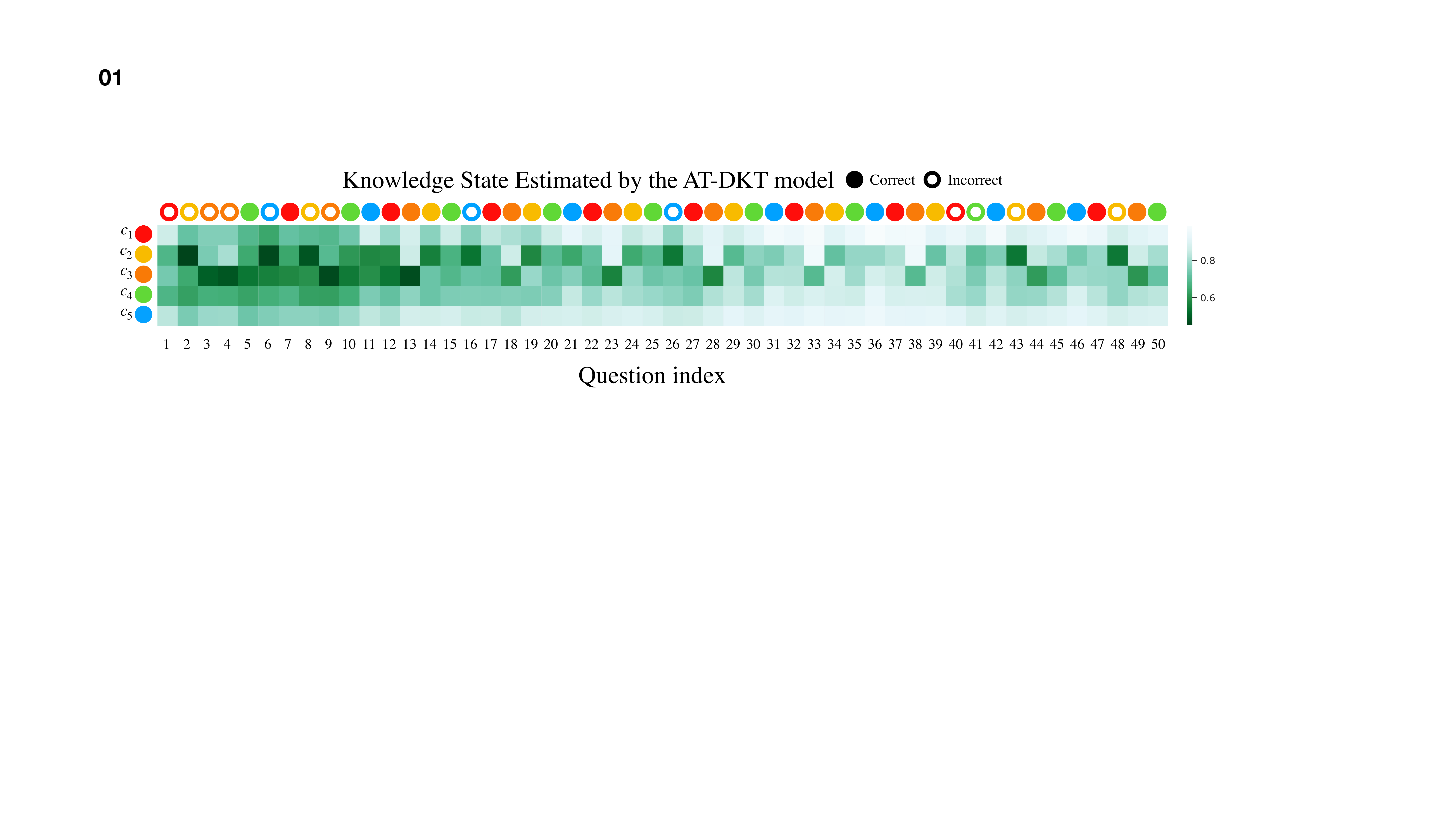} \vspace{-0.3cm}
\caption{An example of knowledge state changes of 5 concepts as the student 3 solves 50 questions from the AL2005 dataset.} 
\label{fig:case_3}
\Description[case]{case}
\end{figure*}

\begin{figure*}[!bpht]
\centering
\includegraphics[width=0.9\textwidth]{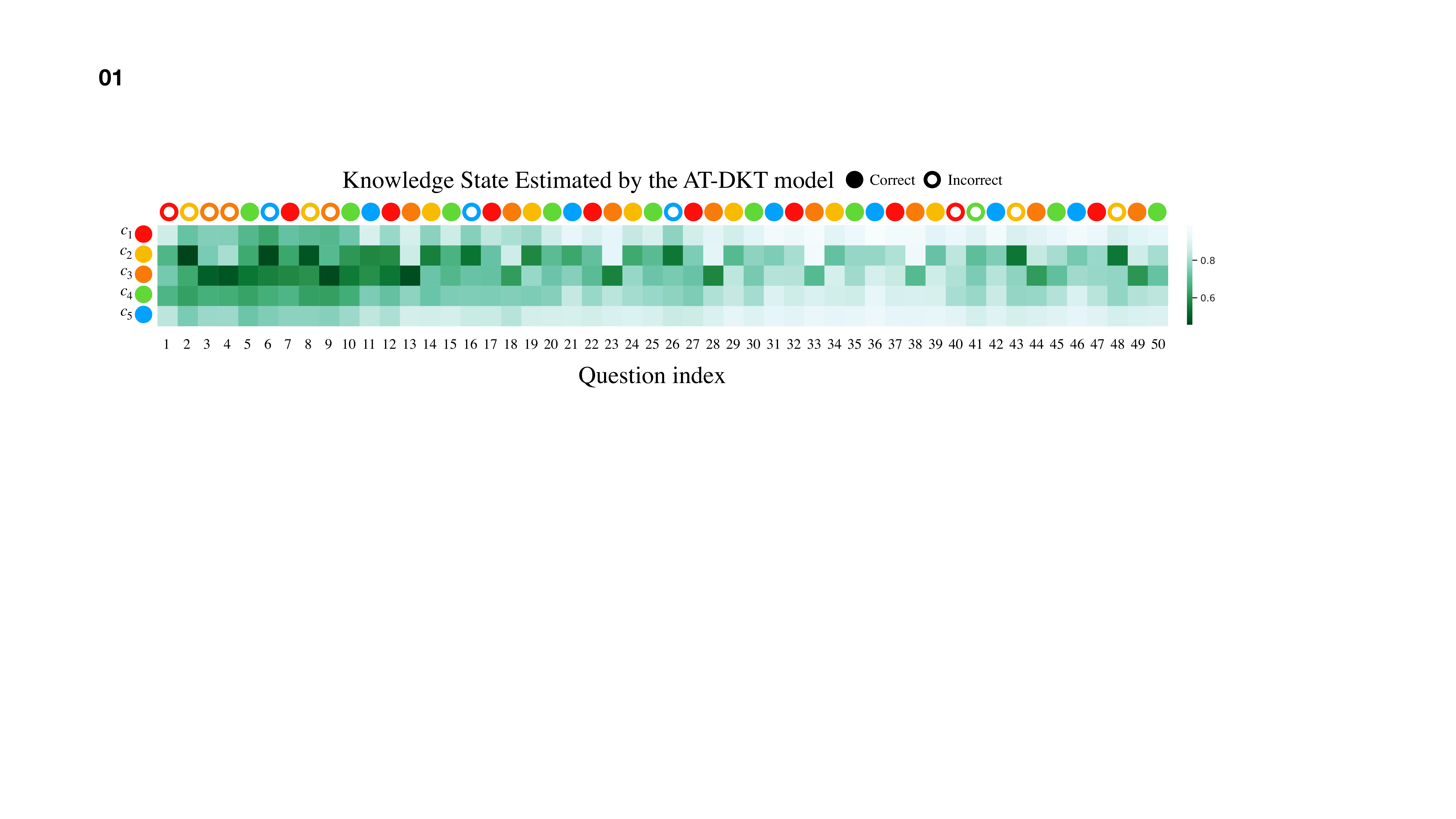} \vspace{-0.3cm}
\caption{An example of knowledge state changes of 5 concepts as the student 4 solves 50 questions from the AL2005 dataset.} 
\label{fig:case_4}
\Description[case]{case}
\end{figure*}

This this section, we show the other three students' prediction outcomes in the AL2005 dataset in Figure \ref{fig:case_2}, \ref{fig:case_3} and \ref{fig:case_4}.

